\documentclass[aps,amsmath,amssymb,prl,twocolumn,showpacs]{revtex4}
\usepackage{bm}
\usepackage{epsfig}

\newcommand{\etal}{{\textit{et al.}}}
\newcommand{\di}{\mathrm{d}}

\newcommand{\br}{{\mathbf r}}

\newcommand{\vs}{{\vec \sigma}}

\newcommand{\spin}{\textrm {s}}
\newcommand{\charge}{\textrm {c}}

\newcommand{\grad}{{\bm \nabla}}

\newcommand{\Ks}{K_\spin}
\newcommand{\Kc}{K_\charge}
\newcommand{\TI}{T_\textrm{I}}

\newcommand{\yv}{y_\textrm{v}}
\newcommand{\yl}{y_\textrm{l}}
\newcommand{\yd}{y_\textrm{d}}

\newcommand{\bQ}{\mathbf{Q}}
\newcommand{\bq}{\mathbf{q}}
\newcommand{\bu}{\mathbf{u}}

\newcommand{\nueff}{\nu_\textrm{eff}}
\newcommand{\lmax}{l_\textrm{max}}

\begin{document}

\title{Non-universal ordering of spin and charge in stripe phases}

\author{Frank Kr\"uger}
\author{Stefan Scheidl}

\affiliation{Institut f\"ur Theoretische Physik, Universit\"at zu
  K\"oln, Z\"ulpicher Str. 77, D-50937 K\"oln, Germany}

\date{\today}

\begin{abstract}
  We study the interplay of topological excitations in stripe phases:
  charge dislocations, charge loops, and spin vortices.  In two
  dimensions these defects interact logarithmically on large
  distances. Using a renormalization-group analysis in the Coulomb gas
  representation of these defects, we calculate the phase diagram and
  the critical properties of the transitions.  Depending on the
  interaction parameters, spin and charge order can disappear at a
  single transition or in a sequence of two transitions (spin-charge
  separation).  These transitions are non-universal with continuously
  varying critical exponents.  We also determine the nature of the
  points where three phases coexist.
\end{abstract}

\maketitle

%%%%%%%%%%%%%%%%%%%%%%%%%%%%%%%%%%%%%%%%%%%%%%%%%%%%%%%%%%%%%%%%%%%%%%%%%%5

High-$T_c$ compounds fascinate not only because of superconductivity
but also because of a variety of concurring orders.  In particular,
theoretical \cite{Schulz89} and experimental \cite{Cheong+91} evidence
has been found for stripes.  Holes which are induced by doping
condense into arrays of parallel rivers in the CuO$_2$ layers.  Within
a layer, each river acts as a boundary between antiferromagnetic
domains with opposite sublattice magnetization.  Thus, stripes are a
combined charge- and spin-density wave.

Based on charge density and magnetization as order parameters, it is
instructive to analyze the interplay of orders in the framework of a
Landau theory \cite{Zachar+98}.  However, in low dimensional
structures fluctuations can be crucial for the nature of phases and of
phase transitions.  In particular, fluctuations play a central role
(i) for spin-charge separation, i.e., the phenomenon that charge order
emerges at higher temperatures than spin order (as observed in
cuprates \cite{Tranquada+95} as well as in nickelates
\cite{Tranquada+96}) and (ii) for the anomalous properties of the
cuprates near optimum doping \cite{Timusk+99}.  To account for
collective low-energy excitations of the electronic system, continuous
deformations of perfect stripe order (spin waves or smooth stripe
displacements) as well as topological defects (such as dislocations,
vortices, or skyrmions) must be considered.  The latter were found to
induce transitions between various liquid-crystal like electronic
phases \cite{Kivelson+98}.  Besides transitions which are related to a
degradation of the charge and spin structure factors, Zaanen {\etal}
\cite{Zaanen+01} have suggested a further transition characterized by
a less accessible, intrinsically topological order.

\begin{figure}[ht]
  \includegraphics[width=0.9\linewidth]{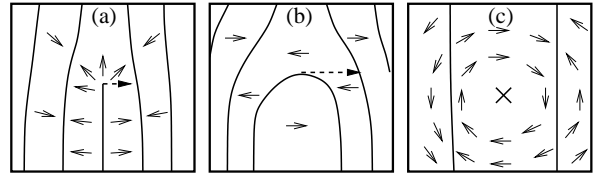}
  \caption{
    Different types of topological defects in the charge density
    (lines) and sublattice magnetization (bold arrows).  (a)
    Combination of a stripe dislocation (dashed arrows correspond with
    the Burger's vectors) and a half-vortex, (b) a charge loop, and
    (c) a vortex.}
  \label{defects} 
\end{figure}

In this Letter we present a paradigmatic model which is amenable to a
largely analytical analysis of the interplay between charge and spin
orders. Motivated by the weakly coupled layered structure of the
materials, we restrict our analysis to two dimensions.  Since the
relevant materials typically have a planar spin anisotropy, the
out-of-plane component of the spins is neglected.  Assuming quantum
fluctuations to be weak in comparison to thermal fluctuations we treat
all degrees of freedom as classical.  In the framework of a
renormalization-group approach we establish the phase diagram and the
nature of the phase transitions.  The latter are driven by three
classes of topological defects (cf.  Fig. \ref{defects}): mixed
defects combining a dislocation in the charge-density wave with a
half-vortex, charge loops (with a Burger's vector of two stripe
spacings) and entire vortices \cite{Zaanen+01}.

\begin{figure}[ht]
 \includegraphics[width=0.9\linewidth]{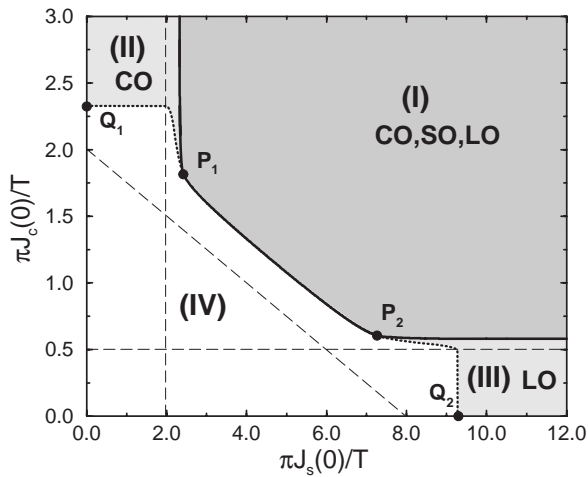}
 \caption{
   Phase diagram.  Thin dashed lines show where defects would become
   relevant for vanishing fugacities.  For finite bare fugacities
   \cite{note:C} the transitions are located at the bold lines.  Phase
   I has no free defects and therefore CO, SO, and LO exist. In phase
   II, vortices proliferate and destroy SO.  In phase III, free loops
   are present, destroying CO and SO.  Eventually, all orders are
   absent in phase IV.}
 \label{phase}
\end{figure}

We identify four different phases (cf. Fig. \ref{phase}), depending on
which types of topological defects proliferate.  In phase I there are
no free defects.  In phase II only vortices, in phase III only charge
loops and in phase IV all types of defects are present.  We
characterize these phases by the range of charge order (CO), spin
order (SO) and collinear order (LO).  The phases are separated by
transition lines along which the correlation functions decay with
\textit{non-universal}, continuously varying exponents.  Slightly
above the temperature $\TI$ of the transition out of phase I, the
short-ranged orders have a correlation length $ \xi \sim \exp\left\{c
  (T-\TI)^{-\nu}\right\} $ with $\nu=\frac 12$ except for the triple
points $P_{1,2}$ (cf. Fig. \ref{phase}) with $\nu= \frac 25$.

To be specific, we assume that the stripes are parallel to the $y$
direction with a spacing $a$.  In the ground state, the charge density
is modulated with a wave vector $\bq=(2\pi/a,0)$.  In the absence of
dislocations, the stripe conformations can be described by a
single-valued displacement field $\bu \equiv (u,0)$ and the domains
between the stripes can be labeled by the function $ \theta(\br)
\equiv \bq \cdot (\br-\bu) + \sum_{m \neq 0} \frac 1{i m} e^{i m \bq
  \cdot (\br-\bu)} $ which increases by $2\pi$ across a stripe.  The
charge density is located at the domain boundary and can be expressed
as $\rho(\br) = \frac \lambda{2\pi} \partial_x \theta(\br)$ where
$\lambda$ denotes the charge per unit length per stripe.  For XY spins
$\vs(\br)\equiv\{\cos \Phi(\br), \sin\Phi(\br)\}$, the magnetization
(including the modulation by the antiferromagnetic order within the
domains and by the anti-phase boundary condition on the stripes) is
captured by the angle $\Phi(\br) = \bQ \cdot \br +
\theta(\br)/2+\phi(\br)$.  Here $\bQ$ is the antiferromagnetic wave
vector and $\phi$ describes the smoothly varying spin Goldstone modes.

In the absence of topological defects low-energy excitations are
wave-like and governed by the Hamiltonian
\begin{eqnarray}
  H_\textrm{wave} = \frac 12 \int d^2r 
  \Big\{ J_{\spin} (\grad \phi)^2 
  + J_{\charge} (2 \pi /a)^2  (\grad u)^2\Big\}
\label{H.wave}
\end{eqnarray}
with spin and charge stiffness constants $J_\alpha$ (with
$\alpha=\spin$, $\charge$, respectively).  The structure factors $
S_\rho(\br-\br') \equiv \langle \delta \rho(\br) \delta \rho(\br')
\rangle $ and $ S_\sigma(\br-\br') \equiv \langle \vs(\br) \cdot
\vs(\br') \rangle $ decay algebraically, $S_\rho(\br) \sim
\cos(\bq\cdot \br) r^{-\eta_\rho}$ with $\eta_\rho=1/(2\pi
K_{\charge})$ and $S_\sigma(\br) \sim \cos[(\bQ+\frac \bq 2) \cdot
\br] r^{-\eta_\sigma} $ with $\eta_\sigma=1/(2\pi K_{\spin}) +1/(8 \pi
K_{\charge})$.  We define the reduced stiffness constants $K_\alpha
\equiv J_\alpha/T$.

In analogy to the Kosterlitz-Thouless (KT) transition
\cite{Kosterlitz+73,Kosterlitz74}, the presence of topological defects
effects a screening of the stiffness constants.  If neutral defect
clusters unbind, these constants are renormalized to zero.  For
suitable values of the stiffness constants charge-loop pairs unbind as
only type of defects (phase III). Then \textit{both} $S_\rho$ and
$S_\sigma$ decay exponentially.  To distinguish this phase from phase
IV where all defects are free, an additional correlation function is
necessary.  For this purpose, Zaanen {\etal}
\cite{Zaanen+97,Zaanen+01} have suggested a highly nonlocal
correlation function involving spin \textit{and} charge.  As a simpler
and probably more physical alternative, we propose
\begin{eqnarray}
  C_\parallel(\br-\br') \equiv 2 \langle (\vs_\br \cdot \vs_{\br'})^2
  \rangle -1 
  \label{ODSO}
\end{eqnarray}
which measures the \textit{spin collinearity} and has the advantage of
being local and being defined entirely in terms of spin variables.
Collinearity is insensitive to continuous stripe displacements and to
loops since the spin ground state remains collinear in the presence of
such excitations.  In the absence of topological defects in the spin
sector (in phase I and III), spin waves lead to an algebraic decay of
collinear order (LO), $C_\parallel(\br) \sim \cos(2 \bQ \cdot \br)
r^{-\eta_\parallel}$ with $\eta_\parallel = 2/(\pi K_{\spin})$.

We now calculate the mutual screening of the defects which, in two
dimensions, interact logarithmically on large distances.  Therefore,
their interactions can be captured by a Coulomb-gas formulation
\begin{equation}
  H_\textrm{top} 
  = \sum_{i\neq j} \sum_{\alpha=\spin,\charge} \pi J_{\alpha} 
  m_{i\alpha} m_{j\alpha}\ln \frac{|\br_i - \br_j|}{a},
\end{equation}
for topological vector charges $(m_{i\spin},m_{i\charge})$ at
positions $\br_i$. Additional core energies of such vector charges
give rise to bare fugacities $y_i \equiv \exp\{-\pi C (K_{\spin}
m_{i\spin}^2 + K_{\charge} m_{i\charge}^2)\}$.  $C$ is a constant of
order unity \cite{note:C}.  Vortices have a vector charge
$(m_{\spin},m_{\charge})=(\pm1,0)$ and a fugacity $\yv = \exp(- \pi C
\Ks)$, and loops are characterized by
$(m_{\spin},m_{\charge})=(0,\pm2)$ and $\yl = \exp(- 4 \pi C \Kc)$.
Dislocations have vector charges $|m_{\charge}|=1$ and
$|m_{\spin}|=\frac{1}{2}$ and fugacities $\yd=\exp\{- \pi C
(\Kc+\Ks/4)\}$.  Defects of higher charges are negligible for the
critical aspects of the phase diagram.
 
We follow the renormalization approach developed by Kosterlitz
\cite{Kosterlitz74} for the usual scalar Coulomb gas and generalized
to vector gases in different contexts \cite{Nelson78,Cardy+82}.  The
charges are assigned a hard-core cutoff $a$ which is increased to
$\tilde a =a e^{\di l}$ under an infinitesimal coarse graining.
Thereby, pairs of vector charges at a distance $a$ annihilate each
other if they have opposite charges, otherwise they recombine to a
single non-vanishing vector charge.  This coarse graining procedure
leads to scale dependent stiffness constants and fugacities, for which
we obtain the flow equations
\begin{subequations}
\begin{eqnarray}
  \frac{\di K_{\spin}^{-1}}{\di l} &=& 2\pi^3(2\yv^2+\yd^2),
  \\
  \frac{\di K_{\charge}^{-1}}{\di l} &=& 8\pi^3(2\yl^2+\yd^2),
  \\
  \frac{\di \yv}{\di l} &=& (2-\pi K_{\spin}) \yv + 2\pi \yd^2,
  \\
  \frac{\di \yl}{\di l} &=& (2-4\pi K_{\charge}) \yl
  +2\pi \yd^2,
  \\
  \frac{\di \yd}{\di l} &=& \left(2-\frac{\pi}{4}
    (K_{\spin}+4K_{\charge})\right) \yd + 2\pi(\yv + \yl) \yd.\quad
\end{eqnarray}
\label{flow}
\end{subequations}
They are invariant under the mapping $(4K_{\charge} , K_{\spin} , \yl,
\yv) \to (K_{\spin} , 4K_{\charge} , \yv , \yl)$. A qualitative
understanding of the phase diagram can be obtained from the rescaling
contributions linear in the fugacities to the flow equations.
Vortices proliferate for $\pi K_{\spin}<2$ and loops for $4\pi
K_{\charge}<2$.  Furthermore, for $\frac{\pi}{4}
(K_{\spin}+4K_{\charge})<2$ dislocations become relevant. The
corresponding borders are shown in figure Fig.  \ref{phase} as dashed
lines.

In order to include the terms quadratic in the fugacities, the flow
equations have to be integrated numerically.  In phase I all
fugacities tend to zero as $l \to \infty$ and $K_{\spin}$ and
$K_{\charge}$ are renormalized to finite values $K_{\spin}(\infty)$
and $K_{\charge}(\infty)$ smaller than the initial ones.  The defects
can be considered as being bound in local clusters of vanishing total
charge.  Their fluctuations enhance the wave excitations such that CO,
SO, and LO are quasi-long ranged: $S_\rho$, $S_\sigma$, and
$C_\parallel$ decay algebraically like in the absence of defects.  The
exponents $\eta$ are given by the expressions given after Eq.
(\ref{H.wave}) if the bare stiffness constants are replaced by the
renormalized ones.  Since the boundary of phase I flows to a line of
fixed points (dashed lines in Fig.  \ref{phase} for zero fugacities)
with continuously varying values of the renormalized stiffness
constants, the transition in non-universal with exponents given in
Tab.  \ref{tab}.

\begin{table}[ht]
  \begin{tabular}{l||lr|lr|lr|l}
    & CO & & SO & & LO & & $\nu$
    \\ \hline\hline
    I/II & $\eta_\rho\le\frac{1}{3}$ & & $\eta_\sigma\le\frac{1}{3}$ &
    * &  $\eta_\parallel=1$ & * &  $\frac{1}{2}$
    \\\hline
    I/III & $\eta_\rho=1$ & * &  $\eta_\sigma\le\frac{1}{3}$ & * &   
    $\eta_\parallel\le\frac{1}{3}$ & &  $\frac{1}{2}$
    \\\hline
    I/IV & $\frac{1}{3}\le\eta_\rho\le1$ & * &  
    $\frac{1}{6}\le\eta_\sigma\le\frac{1}{2}$ & * &
    $\frac{1}{3}\le\eta_\parallel\le1$ & * &  $\frac{1}{2}$
    \\\hline
    II/IV & $\eta_\rho=\frac{1}{4}$ & * &  $\eta_\sigma=\infty$ & & 
    $\eta_\parallel=\infty$ & &  $\frac{1}{2}$ 
    \\\hline
    III/IV & $\eta_\rho=\infty$ & &  $\eta_\sigma=\infty$ & &  
    $\eta_\parallel=\frac{1}{4}$ & * &  $\frac{1}{2}$ 
    \\\hline
    $P_1$ & $\eta_\rho=\frac{1}{3}$ & &  $\eta_\sigma=\frac{1}{3}$ & &
    $\eta_\parallel=1$ & &  $\frac{2}{5}$
    \\\hline
    $P_2$ & $\eta_\rho=1$ & &  $\eta_\sigma=\frac{1}{3}$ & &  
    $\eta_\parallel=\frac{1}{3}$ & &  $\frac{2}{5}$
  \end{tabular}
  \caption{%
    Values of the exponents $\eta,\nu$ at the phase transitions. 
    Orders  that become short-ranged at a transition line ($\eta$ jumps
    to infinity) are marked by an asterisk.
    }% 
\label{tab}
\end{table}

Outside phase I, at least one fugacity increases with the scale. Since
our flow equations are valid only for small fugacities, they can be
evaluated outside the ordered phase only up to a finite scale $l^*$
where some fugacity becomes of the order of unity.  The divergence of
a fugacity drives one or both stiffness constants to zero.  A
divergence of the renormalized exponent $\eta$ then signals
short-ranged order with a finite correlation length, which scales like
$\xi \simeq a e^{l^*}$ sufficiently close to phase I.  Although the
divergence of fugacities hints at the nature of the phases II, III,
and IV, the precise shape of these phases is determined by strong
coupling (large fugacity) regimes beyond the validity of Eqs.
(\ref{flow}).

In phase II, vortices proliferate and the spin stiffness $K_{\spin}$
is renormalized to zero.  This leads to a destruction of the spin and
the collinear order, since the exponents $\eta_\sigma$ and
$\eta_\parallel$ are infinite.  The corresponding correlation
functions decay exponentially, $S_\sigma \sim C_\parallel \sim \exp(-r
/ \xi)$.  Since $K_\spin$ is renormalized to zero, the interaction in
the spin sector effectively breaks down.  Nevertheless, dislocations
and loops are still coupled in the charge sector and can remain bound
for sufficiently large $K_\charge$ \cite{note:strong}.  Then also
charge order remains quasi-long ranged with finite $\eta_\rho$.

In phase III loop pairs unbind and the charge stiffness $K_{\charge}$
is renormalized to zero, leading to $\eta_\sigma=\eta_\rho=\infty$ and
therefore to a short ranged spin and charge order.  The spin order is
destroyed because of the arbitrarily large fluctuations of the domain
walls. Since $K_{\spin}$ is screened only by bound vortices and
dislocations, $\eta_\parallel$ remains finite and quasi-long ranged
collinear order is preserved.

In phase IV the proliferating dislocations obviously render all
correlations short ranged, $S_\sigma \sim S_\rho \sim C_\parallel \sim
\exp(-r / \xi)$.

As pointed out above the precise location of the transitions between
phases II and IV and phases III and IV cannot be obtained by the weak
coupling (low fugacity) flow equations.  Nevertheless, we now present
qualitative arguments for their location, focusing on the example of
the transition II/IV.  In the limit $K_{\spin}(0)=0$ the interaction
in the charge sector is simply switched off.  Nevertheless, loops and
dislocations interact in the charge sector; the former stronger than
the latter.  Therefore, the transition (point $Q_1$ with
$\Kc(\infty)=2/\pi$) is driven by the unbinding of the dislocations.
At this point loops are irrelevant and the transition belongs to the
KT universality class.  When the bare $\Ks$ is increased, it is
renormalized to zero by free vortices and the transition occurs
practically at the same value of $\Kc$ as long as $\Ks \lesssim
2/\pi$.  When $\Ks$ increases to the value where vortices start to
bind, the transition line II/IV reaches the point $P_1$ at a smaller
value of $\Kc$ since dislocations start to be stabilized by the
additional interaction in the spin sector.  The situation for the
transition line between the phases III and IV is similar, in this case
the interaction of the charge components of the mixed defects is
screened by the proliferating charge loops.
  
\begin{figure}[ht]
  \includegraphics[width=0.9\linewidth]{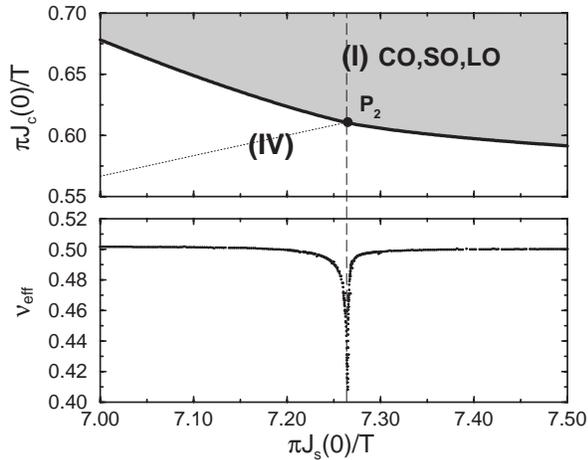}
  \caption{
    Top: magnification of the phase diagram near point $P_2$.  In this
    region, phase III is extremely narrow and masked by the transition
    line to the right of $P_2$.  Bottom: Numerically calculated
    effective exponent $\nueff$ for $\lmax=10^3$ along the boundary of
    phase I.}
  \label{cexp}
\end{figure}

At the triple points $P_{1,2}$ the critical properties are governed by
the balanced competition between two types of defects which can
recombine into each other.  Therefore, these points do not belong to
the universality class of the KT transition.  Due to the symmetry of
the flow equations we concentrate on point $P_2$.  There one can
neglect $\yv$ to a good approximation and the conditions $\yl=\yd$ and
$K_{\spin} = 12K_{\charge}$ (thin dotted line in Fig.  \ref{cexp} top)
are conserved under the flow. In this case the flow equations for
$Y:=2 \pi \sqrt 6 \yd$ and the small variable $x:=\frac{1}{\pi
  K_{\charge}}-2$ read $ \frac{\di x}{\di l} = Y^2 $ and $ \frac{\di
  Y}{\di l} = xY + \frac{1}{\sqrt 6} Y^2.  $ Following Refs.
\cite{Nelson78}, these differential equations can be solved
analytically and result in a critical exponent $\nu=\frac{2}{5}$.

Thus, the exponent $\nu$ is discontinuous along the border of phase I.
While this discontinuity describes the divergence of $\xi$ in the
thermodynamic limit, it is unlikely to be seen in samples of finite
size.  For large but finite systems sizes, $\xi$ appears to diverge
with an effective exponent $\nueff$ which appears to vary continuously
in the range $0.4 \lesssim \nueff \lesssim 0.5$.  We have analyzed the
divergence of $\xi$ from a numerical integration of the flow equations
up to $\lmax=10^3$ which corresponds to astronomically large scales.
The resulting $\nueff$ (cf.  Fig.  \ref{cexp}) shows a notch of a
relatively small width which can become substantially larger for
smaller values of $\lmax$.

From the phase diagram of our model several scenarios of spin-charge
separation are possible.  At low temperatures the stripe phase I is
realized.  (i) For $J_\charge > \frac{3}{4} J_\spin$, there occur
\textit{two distinct transitions} with increasing temperature: first
into phase II (loss of SO and LO), then into phase IV (loss of CO).
This scenario is observed in experiments
\cite{Tranquada+95,Tranquada+96}. (ii) For $\frac{1}{12}J_\spin <
J_\charge < \frac{3}{4} J_\spin$ there is a \textit{single transition}
from phase I to phase IV where all orders disappear simultaneously.
(iii) For $J_\charge < \frac{1}{12} J_\spin$, CO and SO disappear at
the same temperature of the phase transition I/III while LO disappears
only at an even higher temperature at the transition III/IV.  The
relation between $J_\spin$ and $J_\charge$ -- which determines the
scenario -- can be tuned by doping.  With increasing doping, the
shrinking stripe distance should lead to a significant increase of
$J_\charge$, whereas $J_\spin$ should change only weakly since it is
essentially determined by the antiferromagnetic exchange coupling.
Therefore, in principle, the last scenario could be realized and the
transition III/IV should be detectable for example by polarized
neutron scattering.

In conclusion, we have examined a model for coupled spin and charge
order in stripe phases. We have shown that the transition between the
various phases are non-universal.  We have identified collinearity as
a physical quantity that allows to discriminate between phases III and
IV.  Several interesting questions invite to future investigations, in
particular the properties of related quantum critical points at zero
temperature and the influence of disorder (see also \cite{Zachar00})
on the nature of ordering.

The authors benefited from stimulating discussions with D. E.  Feldman
and thank S. Bogner for a critical reading of the manuscript.  This
work was supported by Deutsche Forschungsgemeinschaft SFB 608.


\begin{thebibliography}{16}
\expandafter\ifx\csname natexlab\endcsname\relax\def\natexlab#1{#1}\fi
\expandafter\ifx\csname bibnamefont\endcsname\relax
  \def\bibnamefont#1{#1}\fi
\expandafter\ifx\csname bibfnamefont\endcsname\relax
  \def\bibfnamefont#1{#1}\fi
\expandafter\ifx\csname citenamefont\endcsname\relax
  \def\citenamefont#1{#1}\fi
\expandafter\ifx\csname url\endcsname\relax
  \def\url#1{\texttt{#1}}\fi
\expandafter\ifx\csname urlprefix\endcsname\relax\def\urlprefix{URL }\fi
\providecommand{\bibinfo}[2]{#2}
\providecommand{\eprint}[2][]{\url{#2}}

\bibitem[{\citenamefont{Schulz}(1989)}]{Schulz89}
\bibinfo{author}{\bibfnamefont{H.~J.} \bibnamefont{Schulz}},
  \bibinfo{journal}{J. Physique} \textbf{\bibinfo{volume}{50}},
  \bibinfo{pages}{2833} (\bibinfo{year}{1989});
\bibinfo{author}{\bibfnamefont{J.}~\bibnamefont{Zaanen}} \bibnamefont{and}
  \bibinfo{author}{\bibfnamefont{O.}~\bibnamefont{Gunnarsson}},
  \bibinfo{journal}{Phys. Rev. B} \textbf{\bibinfo{volume}{40}},
  \bibinfo{pages}{7391} (\bibinfo{year}{1989});
\bibinfo{author}{\bibfnamefont{V.~J.} \bibnamefont{Emery}},
  \bibinfo{author}{\bibfnamefont{S.~A.} \bibnamefont{Kivelson}},
  \bibnamefont{and} \bibinfo{author}{\bibfnamefont{H.-Q.} \bibnamefont{Lin}},
  \bibinfo{journal}{Phys. Rev. Lett.} \textbf{\bibinfo{volume}{64}},
  \bibinfo{pages}{475} (\bibinfo{year}{1990}).


\bibitem[{\citenamefont{Cheong et~al.}(1991)\citenamefont{Cheong, Aeppli,
  Mason, Mook, Hayden, Canfield, Fisk, Clausen, and Martinez}}]{Cheong+91}
\bibinfo{author}{\bibfnamefont{S.-W.} \bibnamefont{Cheong}},
  \bibinfo{author}{\bibfnamefont{G.}~\bibnamefont{Aeppli}},
  \bibinfo{author}{\bibfnamefont{T.~E.} \bibnamefont{Mason}},
  \bibinfo{author}{\bibfnamefont{H.}~\bibnamefont{Mook}},
  \bibinfo{author}{\bibfnamefont{S.~M.} \bibnamefont{Hayden}},
  \bibinfo{author}{\bibfnamefont{P.~C.} \bibnamefont{Canfield}},
  \bibinfo{author}{\bibfnamefont{Z.}~\bibnamefont{Fisk}},
  \bibinfo{author}{\bibfnamefont{K.~N.} \bibnamefont{Clausen}},
  \bibnamefont{and} \bibinfo{author}{\bibfnamefont{J.~L.}
  \bibnamefont{Martinez}}, \bibinfo{journal}{Phys. Rev. Lett.}
  \textbf{\bibinfo{volume}{67}}, \bibinfo{pages}{1791} (\bibinfo{year}{1991});
\bibinfo{author}{\bibfnamefont{T.~E.} \bibnamefont{Mason}},
  \bibinfo{author}{\bibfnamefont{G.}~\bibnamefont{Aeppli}}, \bibnamefont{and}
  \bibinfo{author}{\bibfnamefont{H.~A.} \bibnamefont{Mook}},
  \bibinfo{journal}{Phys. Rev. Lett.} \textbf{\bibinfo{volume}{68}},
  \bibinfo{pages}{1414} (\bibinfo{year}{1992});
\bibinfo{author}{\bibfnamefont{S.~M.} \bibnamefont{Hayden}},
  \bibinfo{author}{\bibfnamefont{G.~H.} \bibnamefont{Lander}},
  \bibinfo{author}{\bibfnamefont{J.}~\bibnamefont{Zaretsky}},
  \bibinfo{author}{\bibfnamefont{P.~J.} \bibnamefont{Brown}},
  \bibinfo{author}{\bibfnamefont{C.}~\bibnamefont{Stassis}},
  \bibinfo{author}{\bibfnamefont{P.}~\bibnamefont{Metcalf}}, \bibnamefont{and}
  \bibinfo{author}{\bibfnamefont{J.~M.} \bibnamefont{Honig}},
  \bibinfo{journal}{Phys. Rev. Lett.} \textbf{\bibinfo{volume}{68}},
  \bibinfo{pages}{1061} (\bibinfo{year}{1992});
\bibinfo{author}{\bibfnamefont{C.~H.} \bibnamefont{Chen}},
  \bibinfo{author}{\bibfnamefont{S.-W.} \bibnamefont{Cheong}},
  \bibnamefont{and} \bibinfo{author}{\bibfnamefont{A.~S.}
  \bibnamefont{Cooper}}, \bibinfo{journal}{Phys. Rev. Lett.}
  \textbf{\bibinfo{volume}{71}}, \bibinfo{pages}{2461} (\bibinfo{year}{1993}).

\bibitem[{\citenamefont{Zachar et~al.}(1998)\citenamefont{Zachar, Kivelson, and
  Emery}}]{Zachar+98}
\bibinfo{author}{\bibfnamefont{O.}~\bibnamefont{Zachar}},
  \bibinfo{author}{\bibfnamefont{S.~A.} \bibnamefont{Kivelson}},
  \bibnamefont{and} \bibinfo{author}{\bibfnamefont{V.~J.} \bibnamefont{Emery}},
  \bibinfo{journal}{Phys. Rev. B} \textbf{\bibinfo{volume}{57}},
  \bibinfo{pages}{1422} (\bibinfo{year}{1998}).

\bibitem[{\citenamefont{Tranquada et~al.}(1995)\citenamefont{Tranquada,
  Sternlieb, Axe, Nakamura, and Uchida}}]{Tranquada+95}
\bibinfo{author}{\bibfnamefont{J.~M.} \bibnamefont{Tranquada}},
  \bibinfo{author}{\bibfnamefont{B.~J.} \bibnamefont{Sternlieb}},
  \bibinfo{author}{\bibfnamefont{J.~D.} \bibnamefont{Axe}},
  \bibinfo{author}{\bibfnamefont{Y.}~\bibnamefont{Nakamura}}, \bibnamefont{and}
  \bibinfo{author}{\bibfnamefont{S.}~\bibnamefont{Uchida}},
  \bibinfo{journal}{Nature (London)} \textbf{\bibinfo{volume}{375}},
  \bibinfo{pages}{561} (\bibinfo{year}{1995});
\bibinfo{author}{\bibfnamefont{T.}~\bibnamefont{Niem{\"o}ller}},
  \bibinfo{author}{\bibfnamefont{N.}~\bibnamefont{Ichikawa}},
  \bibinfo{author}{\bibfnamefont{T.}~\bibnamefont{Frello}},
  \bibinfo{author}{\bibfnamefont{H.}~\bibnamefont{H{\"u}nnefeld}},
  \bibinfo{author}{\bibfnamefont{N.~H.} \bibnamefont{Andersen}},
  \bibinfo{author}{\bibfnamefont{S.}~\bibnamefont{Uchida}},
  \bibinfo{author}{\bibfnamefont{J.~R.} \bibnamefont{Schneider}},
  \bibnamefont{and} \bibinfo{author}{\bibfnamefont{J.~M.}
  \bibnamefont{Tranquada}}, \bibinfo{journal}{Eur. Phys. J. B}
  \textbf{\bibinfo{volume}{12}}, \bibinfo{pages}{509} (\bibinfo{year}{1999}).

\bibitem[{\citenamefont{Tranquada et~al.}(1996)\citenamefont{Tranquada,
  Buttrey, and Sachan}}]{Tranquada+96}
\bibinfo{author}{\bibfnamefont{J.~M.} \bibnamefont{Tranquada}},
  \bibinfo{author}{\bibfnamefont{D.~J.} \bibnamefont{Buttrey}},
  \bibnamefont{and} \bibinfo{author}{\bibfnamefont{V.}~\bibnamefont{Sachan}},
  \bibinfo{journal}{Phys. Rev. B} \textbf{\bibinfo{volume}{54}},
  \bibinfo{pages}{12318} (\bibinfo{year}{1996});
\bibinfo{author}{\bibfnamefont{S.-H.} \bibnamefont{Lee}} \bibnamefont{and}
  \bibinfo{author}{\bibfnamefont{S.-W.} \bibnamefont{Cheong}},
  \bibinfo{journal}{Phys. Rev. Lett.} \textbf{\bibinfo{volume}{79}},
  \bibinfo{pages}{2514} (\bibinfo{year}{1997}).

\bibitem[{\citenamefont{Timusk and Statt}(1999)}]{Timusk+99}
\bibinfo{author}{\bibfnamefont{T.}~\bibnamefont{Timusk}} \bibnamefont{and}
  \bibinfo{author}{\bibfnamefont{B.}~\bibnamefont{Statt}},
  \bibinfo{journal}{Rep. Prog. Phys.} \textbf{\bibinfo{volume}{62}},
  \bibinfo{pages}{61} (\bibinfo{year}{1999}).

\bibitem[{\citenamefont{Kivelson et~al.}(1998)\citenamefont{Kivelson, Fradkin,
  and Emery}}]{Kivelson+98}
\bibinfo{author}{\bibfnamefont{S.~A.} \bibnamefont{Kivelson}},
  \bibinfo{author}{\bibfnamefont{E.}~\bibnamefont{Fradkin}}, \bibnamefont{and}
  \bibinfo{author}{\bibfnamefont{V.~J.} \bibnamefont{Emery}},
  \bibinfo{journal}{Nature (London)} \textbf{\bibinfo{volume}{393}},
  \bibinfo{pages}{550} (\bibinfo{year}{1998}).

\bibitem[{\citenamefont{Zaanen et~al.}(2001)\citenamefont{Zaanen, Osman, Krius,
  Nussinov, and Tworzydlo}}]{Zaanen+01}
\bibinfo{author}{\bibfnamefont{J.}~\bibnamefont{Zaanen}},
  \bibinfo{author}{\bibfnamefont{O.~Y.} \bibnamefont{Osman}},
  \bibinfo{author}{\bibfnamefont{H.~V.} \bibnamefont{Krius}},
  \bibinfo{author}{\bibfnamefont{Z.}~\bibnamefont{Nussinov}}, \bibnamefont{and}
  \bibinfo{author}{\bibfnamefont{J.}~\bibnamefont{Tworzydlo}},
  \bibinfo{journal}{Phil. Mag. B} \textbf{\bibinfo{volume}{81}},
  \bibinfo{pages}{1485} (\bibinfo{year}{2001}).

\bibitem[{not({\natexlab{a}})}]{note:C}
\bibinfo{note}{In our numerical calculations, we use bare fugacities with
  $C=\frac{1}{2}\ln(8e^{2\gamma})$ with Euler's constant $\gamma$,
  corresponding to a representation of the model on a square lattice
  \cite{Kosterlitz74}.}

\bibitem[{\citenamefont{Kosterlitz and Thouless}(1973)}]{Kosterlitz+73}
\bibinfo{author}{\bibfnamefont{J.~M.} \bibnamefont{Kosterlitz}}
  \bibnamefont{and} \bibinfo{author}{\bibfnamefont{D.~J.}
  \bibnamefont{Thouless}}, \bibinfo{journal}{J. Phys. C}
  \textbf{\bibinfo{volume}{6}}, \bibinfo{pages}{1181} (\bibinfo{year}{1973}).

\bibitem[{\citenamefont{Kosterlitz}(1974)}]{Kosterlitz74}
\bibinfo{author}{\bibfnamefont{J.~M.} \bibnamefont{Kosterlitz}},
  \bibinfo{journal}{Physica C} \textbf{\bibinfo{volume}{7}},
  \bibinfo{pages}{1046} (\bibinfo{year}{1974}).

\bibitem[{\citenamefont{Zaanen and van Saarloos}(1997)}]{Zaanen+97}
\bibinfo{author}{\bibfnamefont{J.}~\bibnamefont{Zaanen}} \bibnamefont{and}
  \bibinfo{author}{\bibfnamefont{W.}~\bibnamefont{van Saarloos}},
  \bibinfo{journal}{Physica C} \textbf{\bibinfo{volume}{282}},
  \bibinfo{pages}{178} (\bibinfo{year}{1997}).



\bibitem[{\citenamefont{Nelson}(1978)}]{Nelson78}
\bibinfo{author}{\bibfnamefont{D.~R.} \bibnamefont{Nelson}},
  \bibinfo{journal}{Phys. Rev. B} \textbf{\bibinfo{volume}{18}},
  \bibinfo{pages}{2318} (\bibinfo{year}{1978});
\bibinfo{author}{\bibfnamefont{B.~I.} \bibnamefont{Halperin}} \bibnamefont{and}
  \bibinfo{author}{\bibfnamefont{D.~R.} \bibnamefont{Nelson}},
  \bibinfo{journal}{Phys. Rev. Lett.} \textbf{\bibinfo{volume}{41}},
  \bibinfo{pages}{121} (\bibinfo{year}{1978});
\bibinfo{author}{\bibfnamefont{A.~P.} \bibnamefont{Young}},
  \bibinfo{journal}{Phys. Rev. B} \textbf{\bibinfo{volume}{19}},
  \bibinfo{pages}{1855} (\bibinfo{year}{1979}).

\bibitem[{\citenamefont{Cardy and Ostlund}(1982)}]{Cardy+82}
\bibinfo{author}{\bibfnamefont{J.~L.} \bibnamefont{Cardy}} \bibnamefont{and}
  \bibinfo{author}{\bibfnamefont{S.}~\bibnamefont{Ostlund}},
  \bibinfo{journal}{Phys. Rev. B} \textbf{\bibinfo{volume}{25}},
  \bibinfo{pages}{6899} (\bibinfo{year}{1982}).

\bibitem[{not({\natexlab{b}})}]{note:strong}
\bibinfo{note}{In phase II, very close to phase I, the flow equations
  (\ref{flow}) show that $\yv$ becomes of order 1 first. On larger scales, the
  flow equations are no longer reliable. They would suggest that the divergence
  of $\yv$ entails the divergence of $\yd$ and thus the destruction of all
  orders. On physical grounds one expects that the strong coupling limit should
  be described by $\Ks=0$ where the coupling of $\yv$ to the other defects
  becomes meaningless.}

\bibitem[{\citenamefont{Zachar}(2000)}]{Zachar00}
\bibinfo{author}{\bibfnamefont{O.}~\bibnamefont{Zachar}},
  \bibinfo{journal}{Phys. Rev. B} \textbf{\bibinfo{volume}{62}},
  \bibinfo{pages}{13836} (\bibinfo{year}{2000});
\bibinfo{author}{\bibfnamefont{S.}~\bibnamefont{Bogner}} \bibnamefont{and}
  \bibinfo{author}{\bibfnamefont{S.}~\bibnamefont{Scheidl}},
  \bibinfo{journal}{Phys. Rev. B} \textbf{\bibinfo{volume}{64}},
  \bibinfo{pages}{054517} (\bibinfo{year}{2001}).



\end{thebibliography}
\end{document}